\documentclass[11pt]{article}
\usepackage{geometry}                
\usepackage[dvips]{color}
\usepackage{graphicx}
\usepackage{amsmath}
\usepackage{amssymb}
\usepackage{latexsym}
\usepackage{epstopdf}
\usepackage{relsize} 

\usepackage{fancyheadings}

\newcommand{\kinetic}{p_x \pm {i\, p_y}}
\newcommand{\spinup}{\psi_{\scriptscriptstyle\uparrow}}
\newcommand{\spinuphat}{\hat{\psi}_{\scriptscriptstyle\uparrow}}
\newcommand{\spinups}[1]{\psi^#1_{\scriptscriptstyle\uparrow}}
\newcommand{\spindown}{\psi_{\scriptscriptstyle\downarrow}}
\newcommand{\spindownhat}{\hat{\psi}_{\scriptscriptstyle\downarrow}}
\newcommand{\spindowns}[1]{\psi^#1_{\scriptscriptstyle\downarrow}}
\newcommand{\bigt}{\bigtriangleup}

\newcommand{\bsym}{\boldsymbol}
\newcommand{\eit}{e^{i\, E\, t}}
\newcommand{\eitv}{e^{i\, \varepsilon\, t}}
\newcommand{\eitm}[1]{e^{#1 i\, E\, t}}
\newcommand{\eitmm}[1]{e^{#1 i\, \varepsilon\, t}}
\newcommand{\bfr}{{(\bf r)}}
\newcommand{\bfrp}[1]{{({\bf r}^\prime )}}
\newcommand{\bfrr}{{({\bf r} - {\bf r}^\prime)}}
\newcommand{\bfh}{{(\bf k)}}
\newcommand{\bfha}[1]{{(\bf #1 k)}}

\newcommand{\ket}[1]{\left | #1 \right\rangle}
\newcommand{\bra}[1]{\left\langle #1 \right |}

\newcommand{\vbarv}{\Psi^{v\bar{v}}}
\newcommand{\vbarphi}{\Psi^{v\bar{v}}}
\newcommand{\Psihat}{\hat{\Psi}}
\newcommand{\Psihatx}{\hat{\Psi}_{\text{\small cont}}}
\newcommand{\vortexm}{{\bf r} -{\bf R}/2}
\newcommand{\vortexp}{{\bf r} +{\bf R}/2}
\newcommand{\ephase}[1]{e^{i \Omega (#1)}}
\newcommand{\ephasem}[1]{e^{-i \Omega (#1)}}

\newcommand{\intd}{\int d^2 r}

\newcommand{\appsection}[1]{\let\oldthesection\thesection
  \renewcommand{\thesection}{Appendix \oldthesection}
  \section{#1}\let\thesection\oldthesection}

\newcommand{\openone}{I}

\let\a=\alpha \let\b=\beta

\let\s=\sigma   
   
\let\D=\Delta

\def\TT{{\cal T}}

\def\KK{{\cal K}}

\def\rr{{\bf r}}

\title{Quantizing Majorana Fermions in a Superconductor}
\author{C. Chamon$^1$, R. Jackiw$^2$, Y. Nishida$^2$, S.-Y. Pi$^1$, L. Santos$^3$\\[1.5ex]
\small\itshape $^1$ Physics Department, Boston University, Boston MA 02215\\[.5ex]
\small\itshape $^2$ Physics Department, MIT, Cambridge MA 02139\\[.5ex]
\small\itshape $^3$ Physics Department, Harvard University, Cambridge MA 02138}
\date{}                                           

\begin{document}
\maketitle
\thispagestyle{fancy}
\begin{abstract}
A Dirac-type matrix equation governs surface excitations in a
topological insulator in contact with an s-wave superconductor. The
order parameter can be homogenous or vortex valued. In the homogenous
case a winding number can be defined whose non-vanishing value signals
topological effects. A vortex leads to a static, isolated, zero energy
solution. Its mode function is real, and has been called ``Majorana."
Here we demonstrate that the reality/Majorana feature is not confined
to the zero energy mode, but characterizes the full quantum field. In
a four-component description a change of basis for the relevant
matrices renders the Hamiltonian imaginary and the full, space-time
dependent field is real, as is the case for the relativistic Majorana
equation in the Majorana matrix representation. More broadly, we show
that the Majorana quantization procedure is generic to
superconductors, with or without the Dirac structure, and follows from
the constraints of fermionic statistics on the symmetries of
Bogoliubov-de Gennes Hamiltonians. The Hamiltonian can always be
brought to an imaginary form, leading to equations of motion that are
real with quantized real field solutions. Also we examine the Fock
space realization of the zero mode algebra for the Dirac-type
systems. We show that a two-dimensional representation is natural, in
which fermion parity is preserved.
\end{abstract}
\section*{\centerline{\underline{Introduction}}}

Majorana bound states arise as zero energy states in two-dimensional
systems involving superconductors in the presence of
vortices.~\cite{Jackiw:1981ee,Read-Green2000,Ivanov,Fu:2008zzb} These
zero modes have attracted much attention recently, in part because of
the possibility that they can realize ``half-qubits'' within
topological quantum computing schemes~\cite{Ivanov}. The basic idea is
that two far away Majorana bound states, real fermions, can be put
together into a complex fermion acting on a two-dimensional Hilbert
space spanned by the states $|0\rangle$ and $|1\rangle$. Hence, two
Majorana fermions comprise one qubit, which is protected against the
environment if the vortices binding the Majorana fermions are kept far
away from each other.

The first example of a zero mode in a two-dimensional superconductor
was presented in Ref.~\cite{Jackiw:1981ee}. More recently it has been
stated that the proximity effect at the interface between an s-wave
superconductor and the surface of a topological insulator can be
described by a planar Dirac equation~\cite{Fu:2008zzb}, providing a
physical realization of the mathematical structure of
Ref.~\cite{Jackiw:1981ee} Other examples of Majorana bound states
arise in systems with a non-relativistic kinetic term and a $p_\pm
\equiv \kinetic$ interaction with a vortex order parameter ({\it
i.e.}, p-wave superconductors).~\cite{Read-Green2000,Ivanov} These types
of bound states have been the subject of much recent
interest~\cite{Gurarie-Radzihovsky,Tewari-etal,Ghaemi-Wilczek,Bergman-LeHur}. The
focus of the discussions of Majorana fermions in superconductors have
focused thus far on the zero modes.

However, Majorana's original work~\cite{Majorana1937} was actually
quite more general, and did not address a single mode but instead a
whole field. What he showed was that it was possible to construct a
representation of the Dirac equation that admits purely real
solutions. The particles that follow from his construction are their
own anti-particles, and thus necessarily neutral. What was striking
about Majorana's proposal was that these particles were fermions --
bosonic neutral particles represented by real fields are common, pions
and vector bosons, such as photons, being simple examples (see
Ref.~\cite{Wilczek-on-Majorana} for a perspective on Majorana
fermions).

In this paper we look at three issues regarding the quantization of
Majorana fermions, beyond simply the zero modes, in
superconductors. First, we look specifically at the case of Dirac-type
systems describing s-wave induced superconductivity on the surface of
topological insulators. There, we find that the entire $\psi$ field of
the superconductor model (and not merely particular modes) obeys
equations that are analogous to the Majorana equations of particle
physics. The equations of motion for the fields can be brought to a
real form, and the fermionic solutions are real and therefore their
own anti-particles. Indeed, other than the fact that surface states
are 2D, the topological insulator-superconductor system can be brought
to the exactly same form that was discussed in Majorana's original
formulation of real relativistic fermions.

Second, we note various topological features of the Dirac-type model.
We compute the Pontryagin index associated with the ${\bf k}$-space
dispersion, and find it to be $\pm 1/2$, which is an indication of the
existance of zero modes in the presence of vortices. We then present
the Fock space level structures that accommodate an isolated, zero
energy state, which arises in the presence of a vortex. In particular
we show that fermion parity can be preserved, even with a single zero
energy state. As we discussed above the Majorana zero modes are
usually thought of ``half'' qubits, as two of them make up a complex
fermion with a two-dimensional Hilbert space. Here we ruffle this
simple view by quantizing the theory in the infinite plane in the
presence of a single vortex. A sole Majorana zero mode exists, but a
two-dimensional Hilbert space remains. In a finite system, another
zero mode would appear at the edge, which is however absent in the
infinite plane.

Third, we show that the Majorana quantization procedure that we
discuss for the Dirac-type equations describing s-wave induced
superconductivity on the surface of topological insulators does
extend, more broadly, to any superconductor. A description of
Bogoliubov-de Gennes Hamiltonians using Majorana modes has been noted
by Senthil and Fisher~\cite{Senthil-Fisher} for systems where spin
rotational symmetry is broken (classes D and DIII of
Ref.~\cite{Zirnbauer-Altland}). Here we show rather generically that
lack of spin rotation symmetry is not a necessity, and thus classes C
and CI of Ref.~\cite{Zirnbauer-Altland}) also realize Majorana
fermions. All one actually needs is to have fermions, and hence these
results hold for any superconducting system made of half-integer
spin particles, regardless of the size of the spin. What we show is
that the constraints imposed by fermionic statistics on the symmetries
of Bogoliubov-de Gennes Hamiltonians always allow one to bring the
Hamiltonian in the Nambu representation to an imaginary form. In turn,
Schr\"odinger's equation with this imaginary Hamiltonian leads to a
real equation of motion for the fields, as in Majorana's
construction. The real field solutions in the constrained doubled
Nambu space can then be quantized as Majorana fields.

\newpage
\section*{\centerline{\underline{Quantum Structure of the Superconducting Model}}}

Let us start by analyzing the planar Dirac-type systems realized on
the surface of a topological insulator, placed in proximity to an
s-wave superconductor. The Hamiltonian density for the model under
discussion acts on two spatial
dimensions.~\cite{Fu:2008zzb,Jackiw:1981ee} 

\begin{eqnarray}
H=\spinups{\ast} p_-\, \spindown + \spindowns{\ast}\, p_+\, \spinup - \mu (\spinups{\ast}\, \spinup + \spindowns{\ast}\, \spindown) \nonumber\\[.5ex]
+ \bigtriangleup \spinups{\ast} \spindowns{\ast} + \bigtriangleup^\ast \spindown \spinup  \hspace{1.25in}
\label{eq:1}
\end{eqnarray}
Here $\psi_{\scriptscriptstyle \uparrow, \downarrow}$ are electron field amplitudes, $p_\pm \equiv - i\partial_x \pm \partial_y, \mu$ is the chemical potential (which was omitted in the ref.~\cite{Jackiw:1981ee}) and $\bigtriangleup (\mathbf{r})$ is the order parameter that is constant in the homogenous case or  takes a vortex profile in the topologically interesting case: $\bigtriangleup  (\mathbf{r}) = v  (r)\, e^{i \theta}, $ in circular coordinates. Equivalently, in a two-component notation
\begin{eqnarray}
H=\psi_i^{\ast} \left({\boldsymbol \sigma} \cdot{\bf p} -\mu\right)_{ij}\psi_j
+\frac{1}{2}\bigtriangleup \psi_i^{\ast}\;i\sigma^2_{ij}\;\psi_j^{\ast}
-\frac{1}{2}\bigtriangleup^{\ast} \psi_i\;i\sigma^2_{ij}\;\psi_j
\;.
\label{eq:1-ins}
\end{eqnarray}
Now $\psi=
\left(
\begin{array}{c}
\psi_{\scriptscriptstyle \uparrow}\\ \psi_{\scriptscriptstyle \downarrow}
\end{array}
\right)$,
and $\boldsymbol\sigma$ comprises the two Pauli matrices $(\sigma^1,\sigma^2)$. The (2+1)-dimensional equations of motion for \eqref{eq:1}, \eqref{eq:1-ins}
\begin{equation}
\begin{array}{lll}
  i\partial_t\, \spinup&=   &p_- \, \spindown - \mu\, \spinup + \bigtriangleup\, \spindowns{\ast}   \\[.5ex]
  i\partial_t\,  \spindown&=   & p_+\,  \spinup - \mu\, \spindown - \bigtriangleup\, \spinups{\ast}
  \label{eq:2}
\end{array}
\end{equation}
can be presented in two-component matrix notation.
\begin{eqnarray}
i\,\partial_t \psi = \left({\boldsymbol\sigma}\cdot{\bf p} -\mu \right) \psi
+\bigtriangleup \,i\sigma^2\;\psi^{\ast}
\label{eq:2-ins}
\end{eqnarray}
When the chemical potential is absent, and $\triangle$ is constant, the above system is a (2+1)-dimensional version of the (3+1)-dimensional, two component Majorana equation, which in (3+1)-dimensional space-time describes chargeless spin 1/2 fermions with ``Majorana mass'' $|\bigtriangleup|$.~\cite{Brown}

A static solution to \eqref{eq:2}, equivalently \eqref{eq:2-ins}, with a vortex profile for  $\triangle$, can be readily found. It  corresponds to a zero energy mode.  With $f$ and $g$ real in the {\it Ansatz} 
\begin{equation}
\begin{array}{lll}
\spinup  & =  & f (r) \exp\, \{ -i\, \pi/4 - V (r) \}  \\ [.5ex]
 \spindown &=   &g (r) \exp\, \{i (\theta + \pi/4) - V (r)\}   \\ [.5ex]
  &   &   V^\prime (r) \equiv v (r)
  \label{eq:3}
\end{array}
\end{equation}
\eqref{eq:2} reduces to
\begin{equation}
\begin{array}{ccc}
 (r\, g)^\prime & =  & \mu\, r\, f   \\
  f^\prime& =  & -\mu g \ .
  \label{eq:4}
\end{array}
\end{equation}
(Dash signifies $r$ - differentiation). Regular solutions are Bessel functions
\begin{equation}
\begin{array}{lll}
f\, (r)  &=   & N J_0 \, (\mu r)  \\[.5ex]
g \, (r) & =  & N J_1\, (\mu r) 
\label{eq:5}
\end{array}
\end{equation}
with $N$ a real normalization constant.~\cite{footnote1}


While the static, zero energy mode is readily obtained from eq.\! \eqref{eq:2}, for the finite energy modes, we must take account of the fact that $\spinup,\spindown$ mix with their complex conjugates. Therefore, one cannot separate the time dependence with an energy phase. Correspondingly one cannot construct a Hamiltonian energy eigenvalue problem, which is the usual first step in the quantization procedure.

Progress is achieved by doublings the system with a four-component spinor.
\begin{equation}
\Psi =
\left(
\begin{array}{c}
 \spinup \\
  \spindown\\
  \spindowns{\ast}\\
 -\spinups{\ast}
\end{array}
\right) 
=
\left(
\begin{array}{c}
\psi \\[.5ex]
 i \sigma^2 \, \psi^\ast
\end{array}
\right)
\;.
\label{eq:6}
\end{equation}
An extended Hamiltonian density $\mathcal{H}$ leads to equations for $\Psi$, which are just two copies of \eqref{eq:2} or \eqref{eq:2-ins}.
\begin{equation}
\mathcal{H} = \frac{1}{2}\ \Psi^{\ast T} \left(\begin{array}{ccc} \bsym{\sigma}  \cdot \bsym{p} - \mu & & \bigt\\[.5ex]
                                                                                                                 \bigt^\ast & & -\bsym{\sigma} \cdot \bsym{p} + \mu
\end{array}
\right) \Psi \equiv \frac{1}{2}\ \Psi^{\ast T} \, h\, \Psi
\label{eq:7}
\end{equation}
Here $T$ denotes transposition. Because the last two components of $\Psi$ are constrained by their relation to the first two, $\Psi$ satisfies the (pseudo) reality constraint
\begin{equation}
C \Psi^\ast = \Psi
\label{eq:8}
\end{equation}
with $C = C^{-1} = C^\ast =  C^T = C^\dagger \equiv \left(\begin{array}{ccc} 0 & - i\, \sigma^2\\[.5ex]  i\, \sigma^2 & 0 \end{array} \right)$. 

To proceed, one ignores the constraint \eqref{eq:8} on $\Psi$, and works with an unconstrained four-spinor $\Phi = \left(\begin{array}{c} \psi\\ \varphi \end{array} \right)$. Time can now be separated with the usual phase {\it Ansatz}, and the energy eigenvalue spectrum can be found.
\begin{equation}
\begin{array}{c}
h \Phi = i\, \partial_t\, \Phi \ , \ \Phi = \eitm{-}\, \Phi_E   \\[.5ex]
 h \Phi_E = E \, \Phi_E
\;.
 \label{eq:9}
\end{array}
\end{equation}
These are the Bogoliubov-de\! Gennes equations for the superconductor problem. In the particle physics application, the unconstrained four-component equation is just the Dirac equation describing charged spin 1/2 fermions. When the (pseudo) reality constraint is imposed, one is dealing with the four-component version of the Majorana equation.~\cite{Brown}

Observe that $h$ in \eqref{eq:7} possesses the conjugation symmetry
\begin{equation}
C^{-1}\, h \, C = - h^\ast ,
\label{eq:10}
\end{equation}
which has the consequence that to each positive energy eigen mode there corresponds a negative energy mode.
\begin{equation}
C\,  \Phi^\ast_{ + E} = \Phi_{- E}
\label{eq:11}
\end{equation}

A quantum field may now be constructed by superposing the energy eigen modes $\Phi_E$ with appropriate creation and annihilation quantum operators. It is here that we again encounter the Majorana construction: the unconstrained fermion four-spinor $\Phi$ is like a ``Dirac" fermion spinor, governed by a Hamiltonian, which satisfies a conjugation symmetry \eqref{eq:10} that leads to \eqref{eq:11}. Then the spinor $\Psi$, which satisfies the (pseudo) reality constraint \eqref{eq:8}, is like a ``Majorana" spinor, {\it viz.} a ``Dirac" spinor obeying a (pseudo) reality condition.

With the eigen modes one can construct a quantum field $\hat{\Phi}$. It can be an unconstrained ``Dirac" field operator.
\begin{equation}
\begin{array}{ccccc}
\hat{\Phi}  & =  & \sum\limits_{E>0}\, a_E\ \eitm{-}\ \Phi_E &+ &\sum\limits_{E<0}\, b^\dagger_{-E}\, \eitm{-}\, \Phi_E \\[.5ex]
  &  = &\sum\limits_{E>0}\, a_E\, \eitm{-}\, \Phi_E  &+ & \sum\limits_{E>0}\, b^\dagger_E\, \eit \, C \Phi^\ast_E
\end{array}
\label{eq:12}
\end{equation}
Here the $a_E$ operator annihilates positive energy excitations (conduction band) and the  $b^\dagger_E$ operator creates negative energy excitations (valence band). Since $\hat{\Phi}$ is unconstrained, $a$ and $b$ are independent operators.
Their conventional anti-commutators ensure that the unconstrained fields satisfy Dirac anti-commutation relations.
\begin{subequations}\label{eq:13ab}
\begin{align}
\left\{\hat{\Phi}_i \bfr, \hat{\Phi}_j \bfrp{\prime}\right\} \ &= \ 0\label{eq:13a}\\
\left\{\hat{\Phi}_i \bfr, \hat{\Phi}^\dagger_j  \bfrp{\prime}\right\} \ & = \ \delta_{ij} \, \delta \bfrr
\label{eq:13b}
\end{align}
\end{subequations}

For the superconductor/topological insulator system under consideration $\Phi \to \Psi$, and the quantum field $\hat{\Psi}$  satisfies the constraint
\begin{equation}
C_{ij}\, {\hat\Psi^{\dagger}_j} = \hat{\Psi}_i.
\label{eq:13}
\end{equation}
This is achieved by setting $b =a$ in \eqref{eq:12}.
\begin{equation}
\hat{\Psi} = \sum\limits_{E>0}\ \left(a_E\, \eitm{-}\, \Phi_E + a^\dagger_E\ \eit\, C \Phi^\ast_E \right)
\label{eq:14}
\end{equation}
Owing to the constraint \eqref{eq:13} the anti-commutators take a ``Majorana" form.
\begin{subequations}\label{eq:16ab}
\begin{align}
\left\{\hat{\Psi}_i \bfr, \hat{\Psi}_j \bfrp{\prime}\right\} \ &= \ C_{ij} \, \delta \bfrr\label{eq:16a}\\
\left\{\hat{\Psi}_i \bfr, \hat{\Psi}^\dagger_j  \bfrp{\prime}\right\} \ & = \ \delta_{ij} \, \delta \bfrr
\label{eq:16b}
\end{align}
\end{subequations}
These also follow from \eqref{eq:14}, with $a_E, a^\dagger_E$ obeying conventional anti-commutators. We have ignored possible zero-energy states; they will be discussed at length below.

In the final result \eqref{eq:14}, $\hat{\Psi}$ retains the Majorana feature of describing excitations that carry no charge. This is true for the entire quantum field $\hat{\Psi}$, not only for its zero energy modes (if any), which are emphasized in the condensed matter literature. Explicitly we see this by examining the conserved current  that  is constructed with the unconstrained ``Dirac" field $\Phi$.
\begin{equation}
(\rho, {\bf J}) = \left(\Phi^\ast_i\, \Phi_i, \ \Phi^\ast_i \left[\begin{array}{cc}{\boldsymbol \sigma} & 0\\ 0 & -{\boldsymbol \sigma}  \end{array}   \right]_{ij}   \Phi_j \right)
\label{eq:new17}
\end{equation}
When the above is evaluated on the constrained field $\Psi$, all terms vanish. This is to be expected for a Majorana field which carries no charge.

One may also consider a chiral current constructed with the ``Dirac" field $\Phi$.
\begin{equation}
(\rho_5, {\bf J}_5) = \left(\Phi^\ast_i \left[\begin{array}{cc} I&0\\[1ex] 0& -I\end{array} \right]_{ij} \Phi_j\, , \, \Phi^\ast_i \left[\begin{array}{cc} {\boldsymbol \sigma}&0\\[1ex] 0& {\boldsymbol \sigma}\end{array} \right]_{ij} \Phi_j \right)
\label{new18}
\end{equation}
But with non-vanishing $\triangle$ this is not conserved.
\begin{equation}
\frac{\partial}{\partial t}\ \rho_5 +{\boldsymbol \nabla} \cdot {\bf J}_5 = - 2i \, \Phi^\ast_i \left(\begin{array}{cc}0 & \triangle  \\ -\triangle^\ast & 0 \end{array}\right)_{ij} \Phi_j
\label{new19}
\end{equation}
These results persist when the constraint \eqref{eq:13} is imposed on $\Phi \to \Psi$.
\begin{alignat}{3}
(\rho_5 , {\bf J}_5) &\Rightarrow& \ 2 (\psi^{\ast T}\, \psi , \psi^{\ast T}\, {\boldsymbol \sigma}\, \psi)\hspace{.75in}\label{new20}\\[1ex]
\frac{\partial}{\partial t} \ \rho_5 +{\boldsymbol \nabla} \cdot {\bf J}_5 & \Rightarrow& 2\triangle\, \psi^{\ast T}\, \sigma^2\, \psi^\ast + 2\triangle^\ast\, \psi^T \, \sigma^2 \, \psi
\label{new21}
\end{alignat}
Thus no conserved current is present in the superconductor model \eqref{eq:1}.

The Majorana/reality properties are obscured by the representation of the Dirac matrices employed in presenting the $4 \times 4$ Hamiltonian $h$ \eqref{eq:7}. As written, the matrices in $h$ are given in the Weyl representation.
\begin{equation}
 {\boldsymbol \alpha} = \left(\begin{array}{cc}
{\boldsymbol \sigma} & 0\\ 
0 & -{\boldsymbol \sigma}  \end{array}   \right) \qquad 
\beta =  \left(\begin{array}{cc}
0 & I\\ 
I & 0  \end{array} \right) \qquad
\gamma_5 = \left(\begin{array}{cc}
I & 0\\ 
0 & -I  \end{array} \right)
\end{equation}
\begin{equation}
h = {\boldsymbol \alpha \cdot {\bf p}} \, - \mu \, \gamma_5 + \beta\, \triangle_R\, - i \beta \, \gamma_5\, \triangle_I 
\;.
\end{equation}
$\triangle_{R, I}$ are the real and imaginary parts of the order parameter. One may pass to the Majorana representation by conjugating with the unitary matrix
\begin{equation}
V = \left(\begin{array}{cc}
Q_- & Q_+\\ 
Q_+ & -Q_-  \end{array} \right)\;e^{i\pi/4}
\, , 
\qquad Q_\pm \equiv \frac{1}{2}\ \left(1 \pm \sigma^2\right)\, .
\label{eq26}
\end{equation}
Then $h$ becomes
\begin{equation}
V^{-1}\, h\, V = \left(\begin{array}{lr}
-p_y&p_x \sigma^1 + i \triangle_I\\ [1ex]
p_x \sigma^1 - i \triangle_I& p_y  \end{array} \right) \quad + \quad 
\left(\begin{array}{lr}
\mu \sigma^2& -\triangle_R\, \sigma^2\\ [1ex]
-\triangle_R\, \sigma^2& -\mu \sigma^2  \end{array} \right).
\end{equation}
This is manifestly imaginary, and the conjugation matrix $C$  in \eqref{eq:8} becomes the identity, so that the (pseudo) reality constraint on $\hat{\Psi}$ becomes a reality condition.~\cite{footnote2}


\newpage
\section*{\centerline{\underline{Homogenous Order Parameter}}}
For constant $\triangle =m\, e^{i \omega}$, we pass to momentum space with an $e^{i\, {\bf k} \cdot {\bf r}}$ {\it Ansatz} in \eqref{eq:9}. The energy eigenvalue is 
\begin{equation}
E = \pm \sqrt{(k \pm \mu)^2 + m^2} 
\label{eq:15}
\end{equation}
with no correlation among the signs. For fixed $k$ there are two $(\pm \mu)$ positive energy solutions and two negative energy solutions. They become doubly degenerate at $\mu = 0$. The degeneracy occurs because at $\mu = 0$, $h$ commutes with $S =
\left(
\begin{array}{cc}
 0 &  e^{i \omega}\, \sigma^3   \\
 e^{-i \omega}\, \sigma^3  &    0 
\end{array}
\right)$ when the phase $\omega$ of $\triangle$ is constant. The energy \eqref{eq:15} is non-vanishing for all values of the parameters; there is no zero-energy state. 

The operators $a_E, a^\dagger_E$ and the eigen modes $\Phi_E$, which are explicitly presented in Appendix A, are labeled by the momentum $\bf k$ and a further ($+, -$) variety describing the two-fold dependence on $\mu$ of $E_\pm \equiv \sqrt{(k \mp \mu)^2 + m^2}\, \geqslant |m|$.  The quantum operator $\hat{\Psi}$ is constructed as in \eqref{eq:14}, which with notational changes [$a_E \to a_n\,  \bfh; \Phi_E \to \Phi_n  \bfh ; \ \ n = (+, -)]$ reads explicitly 
\begin{equation}
\begin{array}{ccc}
  \Psihat \, (t, {\bf r})& = & \sum\limits_n\, \int  \frac{d^2 k}{(2\pi)^2}\, \Big\{a_n\, \bfh\, e^{-i (E_n t - {\bf k} \cdot {\bf r})}  \Phi_n\, \bfh\\[2ex]
                           &     &+ \, a^\dagger_n \bfh \, e^{i (E_n t - {\bf k} \cdot {\bf r})} C \, \Phi ^\ast_n \, \bfh \Big\}
\label{eq:16}
\end{array}
\end{equation}
with positive energy eigenfunction $\Phi_n \bfh$ carrying energy eigenvalue $E_n$. 
The conjugation condition \eqref{eq:11} now states that $C \Phi^\ast_n \bfh$ is a negative energy solution at $\bfha{-}$. 

Actually we can suppress the lower two components of $\Phi_n \bfh$ in \eqref{eq:16}, because they repeat the information contained in the upper two components, owing to the subsidiary condition \eqref{eq:13}. In this way from the four-component spinors recored in Appendix A, we arrive at a mode expansion for the electron field operators $\hat{\psi}_{\uparrow , \downarrow}$.
\begin{align}
&\hspace{.75in}\hat{\psi} = 
\left(
\begin{array}{ccc}
\spinuphat   \nonumber\\
\spindownhat
\end{array}
\right) =  \nonumber\\[1ex]
&\int \ \frac{d^2 k}{(2\pi)^2}\ \bigg\{\frac{1}{\sqrt{2E_+}} \bigg(a_+ \bfh\, e^{-i (E_+ t- {\bf k} \cdot {\bf r})} \ \frac{\triangle}{\sqrt{E_+ - k + \mu}} \ \ket{+} \nonumber\\[1ex]
&\hspace{.75in} - a^\dagger_+ \bfh\, e^{i (E_+ t- {\bf k} \cdot {\bf r})} \ \sqrt{E_+ - k + \mu} \ e^{-i \varphi} \ket{-} \bigg) \nonumber\\[1ex]
&\hspace{.5in} + \frac{1}{\sqrt{2E_-}}\ \bigg(a_- \bfh \, e^{-i (E_- t- {\bf k} \cdot {\bf r})} \ \frac{\triangle}{\sqrt{E_- + k + \mu}} \ket{-} \nonumber\\[1ex]
&\hspace{.75in} + a^\dagger_- \, \bfh\, e^{i (E_- t- {\bf k} \cdot {\bf r})}\  \sqrt{E_- + k + \mu}\ e^{-i \varphi} \ket{+} \bigg) \bigg\}
\label{eq:17}
\end{align}
Here $\ket{\pm}$ are the 2-component eigenvectors of ${\boldsymbol \sigma \cdot \hat{\bf k}} : \ket{+} \equiv \frac{1}{\sqrt{2}}\ \left(\begin{array}{c} 1\\ e^{i \varphi} \end{array}\right), \ket{-} \equiv \frac{1}{\sqrt{2}} \ \left(\begin{array}{c} 1 \\ -e^{i \varphi}\end{array}\right)$, where $\varphi$ arises as $\hat{k} = (\cos \varphi, \sin \varphi)$. One verifies that \eqref{eq:17} satisfies \eqref{eq:2}. 

The ``Majorana'' character of this expression manifests itself in that the particle annihilation operators $a_\pm$, associated with the positive energy eigenvalues $E_\pm$, are partnered with their Hermitian adjoint creation operators $a_\pm^\dagger$, which are associated with the negative energy $-E_\pm$ modes. By contrast, for a ``Dirac'' field the negative energy modes are associated with the anti-particle creation operators $b_\pm^\dagger$, which anti-commute with $a_\pm,a_\pm^\dagger$. In other words, in the Majorana field operator \eqref{eq:17} the anti-particle (hole) states are identified with the particle states.

\newpage

\section*{\centerline{\underline{Topological numbers}}}

When $\mu$ is absent and $S$ commutes with $h$, we may equivalently work with $h^\prime \equiv S \, h$, which possesses the same eigenvectors as $h$, common with the eigenvectors of $S$. However, $h^\prime$ has the appealing form
\begin{equation}
h^\prime = \Sigma_a  n^a \quad (a = 1, 2, 3).
\label{eq:18-2}
\end{equation}
Here $n^i = k^i \, (i= 1, 2)$ and $n^3$ is $\triangle\, e^{-i \omega} \equiv m$, {\it i.e} the constant phase of $\triangle$ is removed, so $m$ is a real constant, but of indefinite sign. The matrices $\Sigma_a$ 
\begin{equation}
\Sigma_i = \left(\begin{array}{cc}0 & -i e^{i \omega}\, \varepsilon^{ij}\, \sigma^j \\ i e^{-i \omega}\, \varepsilon^{ij}\, \sigma^j & 0\end{array}\right), \Sigma_3 = \left(\begin{array}{cc} \sigma^3 & 0 \\0 & \sigma^3\end{array}\right)
\label{eq:19-1}
\end{equation}
satisfy the SU(2) algebra, as is explicitly recognized after a further unitary transformation.
\begin{equation}
U^{-1}\, \Sigma_a\, U = \left(\begin{array}{cc} \sigma^a & 0 \\0 & \sigma^a\end{array}\right)
\label{eq:20-1}
\end{equation}
\begin{equation}
U \equiv \left(\begin{array}{cc}P_+ & -e^{i\omega} P_- \\ e^{-i\omega}  P_- & P_+\end{array}\right) \, , \, P_\pm \equiv \frac{1}{2} \, (I \pm \sigma_3)
\label{eq:21-1}
\end{equation}

When the Hamiltonian is of the form \eqref{eq:18-2}, we can consider the topological current in momentum space. \cite{Qi:2008ew}

\begin{equation}
K^\mu =  \frac{1}{8\pi}\ \varepsilon^{\mu\alpha\beta}\, \varepsilon_{a b c}\, \hat{n}^a\, \partial_\alpha\, \hat{n}^b\, \partial_\beta\, \hat{n}^c\ \ 
\left(\hat{n} \equiv {\bf n}/|{\bf n}|\right),
\label{eq:22-1}
\end{equation}
and evaluate the topological number by integrating over 2-dimensional $\bf k$-space.
\begin{equation}
{\cal N} = \int d^2 k\, K^0 \bfh = \frac{1}{8\pi}\, \int\, d^2 k \ \frac{m}{(k^2 + m^2)^{3/2}} = \frac{m}{2 |m |}
\label{eq:23-1}
\end{equation}
The non-vanishing answer $\pm 1/2$, depending only on the sign of $m$, is evidence that the model belongs to a topologically non-trivial class. It is also a hint that topologically protected zero modes exist in the presence of a vortex. [Although vortex based zero modes are also present for $\mu \ne 0$, we do not know how to define a winding number in that case.]

We can understand the fractional value for $\mathcal{N}$. The unit vector 
\begin{equation}
\hat{n}^a = (k \cos \varphi, k \sin \varphi, m) / \sqrt{k^2 + m^2}
\label{eq:25new}
\end{equation}
maps $R^{(2)} (\ne S^{(2)}) \ \text{to} \ S^{(2)}$.  When  $k$ begins at $k=0$, $\hat{n}^a$ is at the north or south pole  $(0, 0, \pm 1)$. As $k$ ranges to $\infty$, $\hat{n}^a$ covers a hemisphere (upper or lower) and ends at the equator of $S^{(2)}$. Thus only one half of $S^{(2)}$ is covered.
\newpage

\section*{\centerline{\underline{Single Vortex Order Parameter}}}

When $\triangle$ takes the vortex form, $\triangle \bfr = v (r) e^{i\theta}$, eq. \eqref{eq:9} possesses an isolated zero energy mode
\begin{equation}
\Psi^v_0 = 
\left(
\begin{array}{c}
   \psi^v_0\\[1ex]
   i \sigma^2\, \psi^{v \ast}_0
\end{array}
\right)
\label{eq:19}
\end{equation}
with $\psi^v_0$ determined by \eqref{eq:3} and  \eqref{eq:5}. Note that
\begin{equation}
C \Psi^{v \ast}_0  = \Psi^v_0
\label{eq:20}
\end{equation}
There are also continuum modes. 

The operator field $\hat{\Psi}$ is now given by an expansion like \eqref{eq:14}, except there is an additional contribution due to the zero mode controlled by the operator $A$.
\begin{equation}
\hat{\Psi} \equiv \sum\limits_{E>0}\ \left(a_E\, \eitm{-}\,  \Phi_E + a^\dagger_E \, \eit\, C \Phi^\ast_E \right) + A\, \sqrt{2}\, \Psi^v_0
\label{eq:21}
\end{equation}
(The $\sqrt{2}$ factor will be explained later.) Due to \eqref{eq:13} and \eqref{eq:20}, $A$ is Hermitian $A = A^\dagger$, anti-commutes with ($a_E, a^\dagger_E$) and obeys
\begin{equation}
\left\{A, A\right\} = 2 A^2 = 1.
\label{eq:22}
\end{equation}

The question arises: How is $A$ realized on states? Two possibilities present themselves: two disconnected one-dimensional representations, or one two-dimensional representation.

In the first instance, we take the ground state to be an eigenstate of $A$. The possible eigenvalues are $\pm\, \frac{1}{\sqrt{2}}$, so there are two ground states, $\ket{0+}$ with eigenvalue $+  \frac{1}{\sqrt{2}}$, and $\ket{0-}$  with eigenvalue $-  \frac{1}{\sqrt{2}}$. No local operator connects the two, and the two towers of states built upon them
\[
a^\dagger_E\, a^\dagger_{E\prime}\, a^\dagger_{E\prime\prime} \ldots  \ket{0 \pm}
\]
define two disconnected spaces of states. Moreover, one observes that $A$ has a non-vanishing expectation value $\bra{0\pm} A \ket{0\pm} = \pm \frac{1}{\sqrt{2}}$. Since $A$ is a fermionic operator, fermion parity is lost. \cite{Semenoff:2006zv}

In the second possibility, with a two-dimensional realization, we suppose that the vacuum is doubly degenerate: call one ``bosonic" $\ket{b}$, the other ``fermionic" $\ket{f}$, and $A$ connects the two
\begin{eqnarray}
A \ket{f} &=& \frac{1}{\sqrt{2}}\ \ket{b} \label{eq:23}\\[1ex]
A \ket{b} &=& \frac{1}{\sqrt{2}}\ \ket{f} \label{eq:24}
\;.
\end{eqnarray}
(Phase choice does not loose generality.)

\noindent Again there are two towers of states 
\[
a^\dagger_E\, a^\dagger_{E\prime}\, a^\dagger_{E\prime\prime} \ldots  \ket{f},  \ \
a^\dagger_E\, a^\dagger_{E\prime}\, a^\dagger_{E\prime\prime} \ldots  \ket{b}
\]
but now $A$ connects them. With this realization, fermion parity is preserved when $\ket{b}$ and $\ket{f}$ are taken with opposite fermion parity. Of course since $A$ is Hermitian, it can be diagonalized by the eigenstates. 
\begin{equation}
\begin{array}{ll}
\ket{0+} & = \frac{1}{\sqrt{2}}\ \left(\ket{b} + \ket{f} \right)\\[1ex]
\ket{0-} & = \frac{1}{\sqrt{2}}\ \left(\ket{b} - \ket{f} \right)
\label{eq:25}
\end{array}
\end{equation}
This regains the two states of the two one-dimensional realizations. But the combination $\ket{b} \pm \ket{f}$ violates fermion parity as it superposes states with opposite fermion parity.

There does not seem to be a mathematical way to choose between the two possibilities. But physical arguments favor the fermion parity preserving realization. First of all, there is no reason to abandon fermion parity; if possible it should be preserved since it is a feature of the action. Also arguments against combining states of opposite fermion parity may be given: Since bosons and fermions transform differently under $2\pi$- spatial rotations; the $\ket{0\pm}$ states in \eqref{eq:25} are not rotationally covariant, but transform into each other. [This argument is completely convincing in a (3+1)-dimensional theory. In (2+1) dimensions the anyon possibility clouds the picture, and in (1+1) dimensions the argument cannot be made, because spatial rotations do not occur.] Furthermore, time inversion transformations work differently on bosons and fermions: $T^2$ is $I$ for bosons and $- I$ for spinning fermions. The superposed states \eqref{eq:25} are not invariant under $T^2$, rather they transform into each other. [This argument can be made for (2+1) dimensional models, but in (1+1) dimensions spin is absent so the fermion parity violating option cannot be ruled out. Furthermore, the fermion-boson equivalence of (1+1)-dimensional models obscures the status of fermion-boson mixing. Indeed it is argued within super-symmetry that fermion parity is lost in the presence of solitons in (1+1) dimensions ``due to boundary effects." \cite{Losev:2001uc}
]

[Any argument based on time inversion transformations requires viewing the complex valued vortex configuration as arising from the degrees of freedom of an enlarged model, in which the vortex emerges from the dynamics of the extended model (Abrikosov, Ginzburg, Landau). Otherwise, a vortex background is not $T$-invariant.]

In the next Section we examine the vortex/anti-vortex background and argue that the two-state, two-dimensional, fermion parity preserving realization can be established. The physical picture that emerges is that there are two towers of states, one built on an ``empty" zero energy state $\ket{b}$, the other on the ``filled" zero-energy state $\ket{f}$, and the $A$ operator, which connects the two ``vacua," fills or empties the zero energy state.

\newpage
\section*{\centerline{\underline{Vortex/anti-Vortex Order Parameter}}}
Insight on physical states in the presence of a vortex in a superconductor adjoined to a topological insulator can be gotten by considering a vortex/anti-vortex background. The zero energy mode for an anti-vortex at the origin, $\triangle (r) = v (r) e^{-i \theta}$, is given by
\begin{equation}
\begin{array}{lll}
\spindown &=& N\, J_0\ (\mu\, r) \exp \left\{i \pi/4 - V (r)\right\}\\[1ex]
\spinup &=& N\, J_1\ (\mu\, r) \exp \left\{-i(\theta +  \pi/4) - V(r) \right\} \ .
\end{array}
\label{eq:26}
\end{equation}
To simplify the discussion, we omit the chemical potential and evaluate $V(r) \equiv \int^r\, d r^\prime \, v (r^\prime)$ with the asymptotic form of $ v (r)_{\ \overrightarrow{r \to \infty}}\ m$. Thus the zero-energy mode for the vortex becomes, approximately
\begin{subequations}
\begin{equation}
\psi^v_0 \approx N\, e^{-i\pi/4}\ e^{-m r}
\left(
\begin{array}{c}
1\\
0
\end{array}
\right)
\label{eq:27a}
\end{equation}
while the anti-vortex at ${\bf r} = {\bf R}$, in the same approximation leads to 
\begin{equation}
\psi^{\bar{v}}_0 = N\, e^{i\pi/4}\ e^{-m |{\bf r} - {\bf R}|}
\left(
\begin{array}{c}
0\\
1
\end{array}
\right) .
\end{equation}
\end{subequations}
The corresponding 4-spinors that solve Eq. \eqref{eq:9} at zero energy are
\begin{subequations}
\begin{eqnarray}
\Psi^v_0 &=& 
\left(
\begin{array}{c}
 N\, e^{-i \pi/4}\, e^{-m r}\\[.5ex]
 0 \\[.5ex]
 0\\[.5ex]
 -N\,  e^{i \pi/4}\, e^{-m r} 
\end{array}
\label{eq:28a}
\right)\\[1ex]
\Psi^{\bar{v}}_0 &=& 
\left(
\begin{array}{c}
0\\[.5ex]
 N\, e^{i \pi/4}\, e^{-m |{\bf r} - {\bf R}|}\\[.5ex]
 N \,e^{-i \pi/4}\,  e^{-m |{\bf r} - {\bf R}|}  \\[.5ex]
 0\\[.5ex]
\end{array}
\right).
\label{eq:28b}
\end{eqnarray}
\end{subequations}

Consider now a configuration with a vortex at the origin and an anti-vortex at ${\bf R}$. No zero mode is present in the spectrum of $h$; rather there are two bound states, one with positive, exponentially small energy $\varepsilon \approx e^{- m R}$ and the other with equal magnitude, but opposite sign.

The former, called $\Phi^{v \bar{v}}_\epsilon$, consists of portions localized at the origin (vortex) and at ${\bf r} = {\bf R}$ (anti-vortex). The latter is given by $\Phi^{v \bar{v}}_{-\epsilon} = C \Phi^{v \bar{v}\ast}_{\epsilon}$, and has similar structure. Both contribute unambiguously to the expansion of the quantum field operator $\hat{\Psi}$, the former with an annihilation operator, the latter with a creation operator.
\begin{equation}
\hat{\Psi}  \equiv  \hat{\Psi}_{\text{\small cont}} + a_\epsilon\,  \eitmm{-}\, \Phi^{v \bar{v}}_\epsilon
                 +\,  a^\dagger_\epsilon\, \eitv \, C \Phi^{v \bar{v} \ast}_\epsilon
\label{eq:29}
\end{equation}
The first term on the right is the continuum contribution, as in \eqref{eq:21}. The Fock space spectrum is clear. There is a vacuum state $\ket{\Omega}$ annihilated by $a_\epsilon$
\begin{subequations}\label{eq:30}
\begin{equation}
a_\epsilon\, \ket{\Omega} = 0.
\label{eq:30a}
\end{equation}
A low-lying state is gotten by operating on $\ket{\Omega}$ with $a^\dagger_\epsilon$.
\begin{eqnarray}
a^\dagger_\epsilon \ket{\Omega} &=& \ket{f}\label{eq:30b}\\[1ex]
a_\epsilon\, \ket{f} &=& \ket{\Omega}\label{eq:30c}\\[1ex]
a^\dagger_\epsilon \ket{f} &=& 0\label{eq:30d}
\end{eqnarray}
\end{subequations}
The remaining states, created by $a^\dagger_E$ can be built either on the vacuum $\ket{\Omega}: a^\dagger_E\, a^\dagger_{E\prime}\, a^\dagger_{E\prime\prime} \ldots \ket{\Omega}$, or on the low lying state $\ket{f} : a^\dagger_\epsilon \ket{\Omega}\!\!:  a^\dagger_E\, a^\dagger_{E\prime}\, a^\dagger_{E\prime\prime} \ldots \ket{f}$.

Now let us remove the anti-vortex by passing $R$ to infinity. Both $\vbarv_{\pm \epsilon}$ collapse to their zero-mode limit, $\vbarphi_{\pm \epsilon}\ \, \raisebox{-.35ex}{${\scriptstyle\overrightarrow{ \epsilon \to 0}}$}\ \,\Psi^v_0$, and the expansion \eqref{eq:23} becomes
\begin{equation}
\begin{array}{lll}
\Psihat &=& \Psihatx  + \left(\frac{a_\epsilon + a^\dagger_\epsilon}{\sqrt{2}}\right)\ \sqrt{2}\ \Psi^v_0\\[2ex]
              &=& \Psihatx + A \, \sqrt{2}\ \Psi^v_0
\label{eq:31}
\end{array}
\end{equation}
Moreover the action of $A \equiv \frac{1}{\sqrt{2}}\ (a_\varepsilon + a^\dagger_\varepsilon)= A^\dagger$ may be read off \eqref{eq:30}. Renaming $\ket{\Omega}$ as $\ket{b}$, we find
\begin{equation}
\begin{array}{lll}
\{A, A\} &=& 1\\[1ex]
A \ket{b} &=& \frac{1}{\sqrt{2}}\ \ket{f}\\[1ex]
A \ket{f} &=&  \frac{1}{\sqrt{2}}\ \ket{b}
\label{eq:32}
\end{array}
\end{equation}
and two towers of states are built upon $\ket{b}$ and $\ket{f}$.

In this way we justify the two-dimensional, fermion parity preserving realization of the zero mode algebra in a superconducting/topological insulator system. 

[Note the occurrence  of the factor $\sqrt{2}$ modifying $\Psi^v_0$. This explains its first appearance in eq. \eqref{eq:21}. This factor compensates in the completeness sum for the loss of the anti-vortex wave function.]

Because no explicit solutions in a vortex/anti-vortex background are available, the argument in this Section is qualitative, without explicit formulas. However, one may consider a one-dimensional example with Majorana fermions in the presence of a kink and/or a kink anti-kink pair.\,\cite{Semenoff:2006zv} In that model one can solve equations explicitly and verify the behavior described here for the two-dimensional vortex case. In this way one also establishes that even in one spatial dimension (in the absence of rotation and spin to enforce fermion parity) the two-dimensional realization of the zero mode algebra is appropriate.

In Appendix B we present an approximate determination of the low-energy eigenvalues in the presence of a vortex/anti-vortex pair. The result supports the above qualitative argument: an exponentially small splitting of the zero-energy mode is established.
Also in the Appendix, we study the two vortex background, and find, within the same approximation that no energy splitting occurs; rather two zero modes persist as anticipated by index theorems.

\newpage
\section*{\centerline{\underline{Quantizing Majoranas Fermions in Generic Superconductors}}}


In the preceding sections we showed that Majorana's quantization
prescription of the Dirac equation directly applies to the the full
quantum field describing the proximity effect of an s-wave
superconductor to surface states of a topological insulator. Below we
shall show that Majorana's quantization prescription of real neutral
fermions is rather generic in superconductors, with or without
Dirac-type dispersions. The construction below is possible for any
half-integer spin (fermionic) particle. The reality conditions on the
fermionic fields follow from symmetries of the Bogoliubov-de Gennes
(BdG) Hamiltonian for superconductors constructed in the Nambu basis.

Let us consider a system with fermionic degrees of freedom
$\psi_{\rr,n,s}$ and $\psi^\dagger_{\rr,n,s}$, where $\rr$ labels
position, $n$ the possible flavors (bands, for instance), and $s$ the
spin (half-integer) along a chosen quantization axis. For simplicity,
we shall define an index $\a\equiv (\rr,n,s)$ that encodes all these
degrees of freedom. The Hamiltonian describing superconductivity in
such a  system can be written as
\begin{eqnarray}
{\cal H}&&=
\sum_{\a,\b} 
\psi^\dagger_\a\;H_{\a\b}\;\psi_\b
+\frac{1}{2}
\psi^\dagger_\a\;\D_{\a\b}\;\psi^\dagger_\b
+\frac{1}{2}
\psi_\b\;\D^*_{\a\b}\;\psi_\a
\nonumber\\
&&=
\sum_{\a,\b} 
\frac{1}{2}
\psi^\dagger_\a\;H_{\a\b}\;\psi_\b
-
\frac{1}{2}
\psi_\b\;H_{\a\b}\;\psi^\dagger_\a
+\frac{1}{2}
\psi^\dagger_\a\;\D_{\a\b}\;\psi^\dagger_\b
+\frac{1}{2}
\psi_\b\;\D^*_{\a\b}\;\psi_\a
\nonumber\\
&&=
\sum_{\a,\b} 
\frac{1}{2}
\psi^\dagger_\a\;H_{\a\b}\;\psi_\b
+
\frac{1}{2}
\psi_\a\;\left(-H^T\right)_{\a\b}\;\psi^\dagger_\b
+\frac{1}{2}
\psi^\dagger_\a\;\D_{\a\b}\;\psi^\dagger_\b
+\frac{1}{2}
\psi_\a\;\D^\dagger_{\a\b}\;\psi_\b
\;.
\end{eqnarray}
Defining
\begin{equation}
\Psi=
\left(
\begin{array}{c}
\psi^{\;}\\\psi^\dagger
\end{array}
\right)
\end{equation}
we can write
\begin{equation}
{\cal H}=\frac{1}{2}\:
\Psi^\dagger
\left(
\begin{array}{cc}
H&\D\\
\D^\dagger&-H^*
\end{array}
\right)
\Psi
\equiv
\Psi^\dagger \;h\;\Psi
\;.
\end{equation}
That $H^T=H^*$ follows from $H=H^\dagger$. Notice that $\D=-\D^T$ is
enforced because of fermionic statistics, and consequently
we can also write
\begin{equation}
\label{eq:h-def}
h=\frac{1}{2}
\left(
\begin{array}{cc}
H&\D\\
-\D^*&-H^*
\end{array}
\right)
\;.
\end{equation}
Let us define
\begin{equation}
C=
\left(
\begin{array}{cc}
0&\openone\\
\openone&0
\end{array}
\right)\;,
\end{equation}
so that $C=C^T=C^*=C^\dagger=C^{-1}$. The operators $\Psi$ must satisfy
the constraint
\begin{equation}
C_{ab}\;\Psi_b^\dagger=\Psi_a
\;,
\end{equation}
where the index $a\equiv (\a,p)$, with $p=\pm$ the Nambu grading
($\Psi_{\a,+}=\psi_\a$ and $\Psi_{\a,-}=\psi^\dagger_\a$). The
fermionic commutation relations of the fields $\psi,\psi^\dagger$
translate into
\begin{equation}
\{\Psi_a,\Psi_b\}=C_{ab}
\quad
{\rm and}
\quad
\{\Psi_a,\Psi^\dagger_b\}=\delta_{ab}
\;.
\end{equation}

\subsection*{\centerline{Conjugation symmetry}}

One can easily check that {\it any}
 BdG-type $h$ as in
Eq.~(\ref{eq:h-def}) possesses the following conjugation symmetry,
\begin{equation}
-h^*=C^*\,h\,C
\;.
\end{equation}
We stress that fermionic statistics underlies this result, as it is
the reason for the minus signs and the complex conjugation in both
terms in the second row of Eq.~(\ref{eq:h-def}).

It follows from this symmetry that positive and negative eigen modes
of $h$ are paired:
\begin{equation}
h\Phi_E=E\,\Phi_E
\quad\Rightarrow\quad
h\left(C\,\Phi^*_E\right)=-E\left(C\,\Phi^*_E\right)
\;,
\end{equation}
or equivalently
\begin{equation}
C\,\Phi^*_{+E}=\Phi_{-E}
\;.
\end{equation}

\subsection*{\centerline{Generic Majorana basis and its real equation of motion}}

Consider a unitary transformation $V$, under which
\begin{equation}
h\to \tilde h=V \,h\, V^\dagger
\,.
\end{equation}
It follows that
\begin{eqnarray}
-{\tilde h}^*
=
V^*\left(-h^*\right){V^*}^\dagger
&=&
V^*\,\left(C^*\,h\,C\right)\,{V^*}^\dagger
\nonumber\\
&=&
V^*\,C^*\,V^\dagger\,V\,h\,V^\dagger\,V\,C\,{V^*}^\dagger
\nonumber\\
&=&
\left(V\,C\,V^T\right)^*\,{\tilde h}\,\left(V\,C\,V^T\right)
\nonumber\\
&=&
{\tilde C}^*\,{\tilde h}\,{\tilde C}
\;,
\end{eqnarray}
so the transformation law of $C$ is
\begin{equation}
C\to\tilde C=V\,C\,V^T
\;.
\end{equation}
(Notice that $\tilde C \tilde C^*=V\,C\,V^T
V^*\,C^*\,V^\dagger=\openone$, so $\tilde C^{-1}=\tilde C^*$ still.)

We will construct below a unitary matrix $V$ such that $\tilde
C=\openone$. This basis is the appropriate Majorana representation for
the generic superconducting system of half-integer spin particles (for
any number of flavors). In this basis, one has $\tilde h=-\tilde h^*$,
so that $\tilde h$ is imaginary, or equivalently $i \tilde h$ is
real. It follows from Schr\"odinger's equation that 
\begin{equation}
\left(\partial_t +i\tilde h\right)\tilde \Psi=0
\;,
\end{equation}
so the equation of motion for the field is purely real and thus admits
purely real solutions. Notice that this path mirrors Majorana's
formulation of the Dirac equation for spin 1/2 particles (he
constructed a purely imaginary representation of the Dirac matrices,
obtaining an equation of motion that was real).

Notice that in this basis the commutation relations become
\begin{equation}
\{\tilde \Psi_a,\tilde \Psi_b\}=\tilde C_{ab}=\delta_{ab}
\quad
{\rm and}
\quad
\{\tilde \Psi_a,\tilde \Psi^\dagger_b\}=\delta_{ab}
\;,
\end{equation}
corresponding to real fermions
\begin{equation}
\tilde \Psi_a=\tilde \Psi^\dagger_a
\;.
\end{equation}

\subsubsection*{\centerline{Construction of $V$}}

The appropriate unitary transformation $V$ which makes $\tilde C=\openone$
is constructed as follows. Because of fermionic statistics, the
time-reversal operator $\Theta$ squares to $-1$. One can write
$\Theta=\TT\,\KK$, where $\KK$ is complex conjugation and $\TT=e^{i\pi
S^y}$, with $S^y$ the y-component of the angular momentum operator (in
a representation such that $S^y$ is a purely imaginary matrix). $\TT$
is a real anti-symmetric matrix ($\TT=\TT^*$ and $\TT^T=-\TT$), with
$\TT^2=e^{i2\pi S^y}=-\openone_{\rm spin\times flavor}$ when spin is
half-integer. For instance, for spin 1/2 particles $\TT=i\s^2$.

Consider the following transformation.
\begin{equation}
V = \left(\begin{array}{cc}
Q_- & -i Q_+\\ 
i Q_+ & -Q_- \\
\end{array} \right)\;e^{i\pi/4}\\
\, , 
\qquad Q_\pm \equiv \frac{1}{2}\left(1\mp i\TT\right)
\end{equation}
[compare with eq. \eqref{eq26}.] Notice that $Q_\pm $ are projectors ($Q_\pm^2=Q_\pm$), that
$Q_\pm^\dagger=Q_\pm$, and that $Q_+^2+Q_-^2=\openone_{\rm spin\times
flavor}$ and $Q_+Q_-=Q_-Q_+=0$. Also notice that
$Q_\pm^*=Q_\pm^T=Q_\mp$. One can then easily check that the above defined $V$ is such that
\begin{eqnarray}
{\tilde C}
=
V\,C\,V^T
=
\openone
\;.
\end{eqnarray}

\newpage
\section*{\centerline{\underline{Summary}}}

In this paper we studied mainly three issues regarding the
quantization of Majorana fermions in superconductors, following
closely Majorana's original definitions, and looked beyond just the
Majorana zero energy modes that are bound to topological defects such
as vortices.

We started by analyzing the specific case of Dirac-type systems
describing s-wave induced superconductivity on the surface of
topological insulators. We showed that the entire $\psi$ field of the
superconductor model (and not merely particular modes) obeys equations
that are analogous to the Majorana equations of particle physics.

We then analyzed the quantization of the theory in the presence of
vortices. We showed that fermion parity can be preserved, even with a
single zero energy state. This quantization scheme shows that one can
obtain a two-dimensional Hilbert in the presence of a single vortex in
an infinite plane, presenting a case where each Majorana fermions can
be, when present in odd numbers, more than ``half'' a qubit.

Finally, we showed that the Majorana quantization procedure that we
discussed for the Dirac-type equations describing s-wave induced
superconductivity on the surface of topological insulators does
extend, more broadly, to any superconductor. The constraints imposed
by fermionic statistics on the symmetries of Bogoliubov-de Gennes
Hamiltonians are sufficient to allow real field solutions in the
constrained doubled Nambu space that can then be quantized as Majorana
fields. This results follows simply from fermionic statistics plus
superconductivity, irrespectively of the presence or absence of any
other symmetries in the problem, such as spin rotation invariance or
time-reversal symmetry.

\section*{Acknowledgment}
We thank V. Sanz, who participated in an early stage of this investigation, for useful discussions. N. Iqbal and D. Park evaluated numerically the integral in (\ref{app11}, \ref{app12b}). This research is supported by DOE grants  DEF-06ER46316 (CC) -05ER41360 (RJ) and -91ER40676 (S-Y P), by the MIT Pappalardo Fellowship (YN) and by a Harvard Teaching Fellowship (LS).



\newpage
\numberwithin{equation}{section}
\appendix
\appsection{}
We present the 4-component, positive energy solutions to \eqref{eq:9}. The eigenvalues
\begin{equation*}
E_\pm = \sqrt{(k\mp\mu)^2 + |\triangle|^2}
\end{equation*}
are associated with the eigenvectors
\begin{equation*}
\Phi_+ \bfh = \frac{1}{\sqrt{E_+}} \left(\begin{array}{c}
     									\frac{\triangle}{\sqrt{E_+ - k +\mu}}\ \ket{+}\\[3ex]
									{\scriptstyle \sqrt{E_+ - k +\mu}} \ \ket{+}
							  \end{array}
					              \right)
\end{equation*}
\begin{equation*}
\Phi_- \bfh = \frac{1}{\sqrt{E_-}} \left(\begin{array}{c}
     									\frac{\triangle}{\sqrt{E_- + k +\mu}}\ \ket{-}\\[3ex]
									{\scriptstyle \sqrt{E_- + k +\mu}} \ \ket{-}
							  \end{array}
					              \right).
\end{equation*}
The negative energy spinors are given by
$C \Phi^\ast_\pm \ (-{\bf k}).$ $\ket{\pm}$ are defined in the text.

\appsection{}
We study the low-lying energy levels of the Dirac-type Hamiltonian $h$ in \eqref{eq:7} with $\mu$ set to zero and $\triangle$ chosen first in an approximate vortex/anti-vortex profile,
\begin{equation}
\triangle_{v \bar{v}} = m \, e^{i\Omega ({\bf r} - {\bf R}/2)} \, e ^{-i \Omega ({\bf r} + {\bf R}/2)}
\label{app1}
\end{equation}
and then similarly with two vortices.
\begin{equation}
\triangle_{v v} = m \, e^{i\Omega ({\bf r} - {\bf R}/2)} \, e ^{i \Omega ({\bf r} + {\bf R}/2)}
\label{app2}
\end{equation}
Here $\Omega$ is the argument of the appropriate vector,
\[
\begin{array}{ccc}
\ephase{\bf r}  & \equiv  & \frac{x + i y}{r} = e ^{i \theta}   \\[1ex]
 \ephase{\pm {\bf R}} &  \equiv & \pm\  \frac{(X + i Y)}{R} = \pm\, e^{i\Theta}
\end{array}
\]
with $x = r \cos \theta, y = r \sin \theta, X = R \cos \Theta, Y = R \sin \Theta$. One vertex is located at ${\bf r} \approx {\bf R}/2$, the anti-vortex or the second vortex at ${\bf r} \approx -{\bf R}/2$. \cite{pxmodel}
\subsection{Vortex/anti-Vortex}
Near the vortex at ${\bf r} \approx {\bf R}/2$ the order parameter $\triangle_{v \bar{v}}$ is approximated by
\begin{equation}
\begin{array}{ccc}
\triangle_{v \bar{v}} \to \triangle_v &=& m\, \ephase{\vortexm}\, \ephasem{\bf R}\\[1ex]
                              			    & =& m\, \ephase{\vortexm}\, e^{-i \Theta}.
\label{app3}
\end{array}
\end{equation}
The zero mode in the presence of $\triangle_v$ differs from \eqref{eq:28a} by a phase, due to the additional phase $e^{i \Theta}$ in $\triangle_v$. Also the location is shifted by ${\bf R}/2$.
\begin{equation}
\psi^v_0 = \left( \begin{array}{c} 
			v\\
			0\\
			0\\
			-v^\ast
	                   \end{array}			
\right),\quad v \equiv \frac{m}{\sqrt{\pi}} \ e^{-i(\pi/4 + \Theta/2)}\, e^{-m |\vortexm|}
\label{app4}
\end{equation}
Similarly, with the order parameter near the anti-vortex at ${\bf r} \approx -{\bf R}/2$ taken as
\begin{equation}
\begin{array}{ccc}
\triangle_{\bar{v} v} \to \triangle_{\bar{v}} &=& m\, \ephase{-\bf R}\, \ephasem{\vortexp}\\[1ex]
                              			    & =& - m\, e^{i \Theta} \ephasem{\vortexp}\, 
\label{app5}
\end{array}
\end{equation}
the zero mode solution replacing \eqref{eq:28b} reads
\begin{equation}
\psi^{\bar{v}}_0 = \left( \begin{array}{c} 
			0\\
			\bar{v}\\
			\bar{v}^\ast\\
			0
	                   \end{array}			
\right), \quad \bar{v} \equiv \frac{m}{\sqrt{\pi}} \ e^{i\left(\frac{3\pi}{4} + \Theta/2 \right)} \ e^{-m |\vortexp|}\, .
\label{app6}
\end{equation}

Next we evaluate the matrix element between $\psi^v_0$ and $\psi^{\bar{v}}_0$ of $h$ in \eqref{eq:7}, with $\mu=0$ and order parameters as in \eqref{app1}. We find that the diagonal matrix elements vanish $\bra{v}h\ket{v} = \bra{\bar{v}}h\ket{\bar{v}} = 0$. The energy shift  $\triangle\, E$,  determined by the off-diagonal elements of $h$, is 
\begin{eqnarray}
\triangle\, E = \pm |\bra{\bar{v}}h\ket{v}| \hspace{2in}
\label{app7} \\[1ex]
\bra{\bar{v}}h\ket{v} = \bra{v}h\ket{\bar{v}}^\ast = \int d^2 r \left(\bar{v}^\ast p_+  v + \bar{v} \, \triangle^\ast_{v\bar{v}}\, v\right) - h. c.
\label{app8}
\end{eqnarray}

The evaluation of the first integrand proceeds by recalling that $p_+ v = \triangle_v\, (\vortexm)\, v^\ast$ and yields, after a shift of integration variable by ${\bf R}/2$,
\begin{subequations}\label{app9}
\begin{equation}
\begin{array}{l}
\mathlarger{\int} d^2 r \, \bar{v}^\ast \, p_+ v = - i \frac{m^3}{\pi}\ \mathlarger{\int}^\infty_0\, r\, d\, r \, e^{-mr}\, \mathlarger{\int}^\pi_{-\pi}\, d \theta\, e^{i (\theta -\Theta)} \\[1ex]
\hspace{1in} exp -m [r^2 + R^2 + 2 r R \cos (\theta - \Theta)]^{1/2}. 
\end{array}
\label{app9a}
\end{equation}
A further shift of $\theta$ by $\Theta$ leaves
\begin{equation}
\begin{array}{c}
\mathlarger{\int} d^2 r \, \bar{v}^\ast \, p_{\scriptstyle +} v = - i \frac{2m^3}{\pi}\ \mathlarger{\int}^\infty_0\, r\, d\, r \, e^{-mr}\, \mathlarger{\int}^\pi_{0} d \theta\, \cos \, \theta \, e^{- m D}\ . \\[2ex]
D \equiv \sqrt{r^2 + R^2 + 2 R\, r \, \cos \theta} 
\label{app9b}
\end{array}
\end{equation}
\end{subequations}

For the second integrand, a similar shift, first by ${\bf R}/2$ and then by $\Theta$ gives 
\begin{subequations}\label{app10}
\begin{equation}
\intd \, \bar{v}\, \triangle^\ast_{v \bar{v}}\, v = i \tfrac{2 m^3}{\pi}\ \int^\infty_0\, r\, d\, r \, e^{-mr}\, \int^\pi_{0} d \theta\ \frac{r + R \cos \theta}{D}\ e^{-m D} \ .
\label{app10a}
\end{equation}
We notice that the $\theta$ integrand may also be presented as $-\frac{1}{m}\ \frac{\partial}{\partial r} \ e^{- m D}$, thereby transforming \eqref{app10a} after an integration by parts into
\begin{equation}
\intd \, \bar{v}\, \triangle^\ast_{v \bar{v}}\, v = i \tfrac{2 m^2}{\pi}\ \int^\infty_0\, d r ( 1 - r m) e^{- mr} \int^\pi_0 d \theta \, e^{- m D} \ .
\label{app10b}
\end{equation}
\end{subequations}
Thus we find that
\begin{equation}
\begin{array}{l}
\hspace{.5in}\bra{\bar{v}}h\ket{v} = i \epsilon \\[1.5ex]
\epsilon = 4 m^2 \mathlarger{\int}^\infty_0\, d r \, e^{- mr} \ \frac{1}{\pi}\, \mathlarger{\int}^\pi_0\, d \theta [1 - r m ( 1 + \cos \theta)]\ e^{- m D}
\end{array}
\label{app11}
\end{equation}
and the energy is shifted from zero by $\pm\ \epsilon$.



Numerical integration of \eqref{app11} at large $R$ yields a result consistent with
\begin{equation}
\epsilon\ \raisebox{-.75ex}{$\scriptstyle \overrightarrow{R \to \infty}$} 
\sqrt{\frac{8}{\pi}}\;m\; (m R)^{1/2} \;e^{-m R}\, .
\label{app12b}
\end{equation}
This may be derived analytically with the following argument. We replace the upper limit ($\infty$) of the $r$ integral by $R$ and approximate $D$ by $R + r \cos \theta$. The $\theta$ integral now leads to modified Bessel functions $I_0$ and $I_1$, and we keep only their large argument, exponential asymptote. The remaining $r$ integral yields \eqref{app12b}.



\subsection{Two Vortices}
The order parameter \eqref{app2} describing two vortices located at ${\bf r} \approx \pm\, {\bf R}/2$ reduces at  ${\bf r} \approx  {\bf R}/2$ to
\begin{subequations}\label{app12}
\begin{equation}
\begin{array}{ccc}
\triangle_{v v} \to \triangle_{v +} &=& m\, \ephase{\vortexm}\, \ephase{\bf R}\\[1ex]
                              			    & =& m\, \ephase{\vortexm}\, e^{i \Theta}
\label{app12a}
\end{array}
\end{equation}
while the one at ${\bf r} \approx - {\bf R}/2$ becomes
\begin{equation}
\begin{array}{ccc}
\triangle_{v v} \to \triangle_{v-} &=& m\, \ephase{-\bf R}\, \ephase{\vortexp}\\[1ex]
                              			    & =& - m\, e^{i \Theta} \ephase{\vortexp}\, .
\end{array}
\end{equation}
\end{subequations}
The corresponding zero modes differ by phases from the vortex solution \eqref{eq:28a} or \eqref{app4} but they retain their spinor structure.
\[
\psi^{v+}_0 = \left( \begin{array}{c} 
			v_+\\
			0\\
			0\\
			-v^\ast_+
	                   \end{array}			
\right)
\]
\begin{equation}
\psi^{v-}_0 = \left( \begin{array}{c} 
			v_-\\
			0\\
			0\\
			-v^\ast_-
	                   \end{array}			
\right)
\label{app13}
\end{equation}
The explicit expressions for $v_+$ and $v_-$ are not needed, because the above form of the spinors guarantees that all matrix elements of $h$ vanish. Thus, within our approximation, the 2-vortex background retains its two zero modes. This is to be expected because asymptotically such a configuration is indistinguishable from a double vortex,
\begin{equation}
\triangle_{v v}\ \raisebox{-.5ex}{$\scriptstyle \overrightarrow{r + \infty}$} \ m\,  \ephase{r}\, \ephase{r} = m\, e^{2\, i\, \theta} \, ,
\label{app14}
\end{equation}
and a double vortex possesses two zero modes.\,\cite{Jackiw:1981ee} 

\end{document}